\documentclass[amsmath,prd,nofootinbib,floatfix,12pt,]{revtex4}

\usepackage{epsfig}
\allowdisplaybreaks
\usepackage{setspace,graphicx}
\usepackage[usenames]{color}
\usepackage{subfigure}
\usepackage{epstopdf}

\def\beqa{\begin{eqnarray}}
\def\eeqa{\end{eqnarray}}
\def\beq{\begin{equation}}
\def\eeq{\end{equation}}
\def\f{\frac}
\def\al{\alpha}
\def\AP0#1{\Pi'_{\rm #1}(0)}
\def\A0#1{\Pi_{\rm #1}(0)}
\def\bi{\bibitem}
\begin{document}

\title{The influence of supersymmetric quirk particles on the W mass increment and the muon g-2 anomaly}
\author{Guo-Li Liu}
\affiliation{School of Physics, Zhengzhou University, Zhengzhou 450000, P. R. China}
\begin{abstract}
In quirk assisted Standard Model, the couplings between the exotic particles "quirks" and the gauge bosons
may contribute to the W mass and muon $g-2$ anomaly reported by FermiLab.
We calculate the contributions from supersymmetric quirk particles as an example.
Imposing the theoretical constraints, we found that the CDF II $W$-boson mass increment 
 constrains strictly the mixing and coupling parameters and the quirk mass $m_F$,
 while muon $g-2$ anomaly can not be explained by the exotic particles with their large masses.
\end{abstract}


\maketitle
\baselineskip=14.9pt
\tableofcontents
\vfill\eject
\baselineskip=14.9pt

\setcounter{footnote}{1}
\setcounter{page}{2}
\setcounter{figure}{0}
\setcounter{table}{0}
\section{Introduction}

 Since it contains the key information of EWSB, precision measurement of W boson mass
can provide a stringent test of the SM, and constrain various new physics models.
Recently, CDF II collaborators at the Fermilab Tevatron collider~\cite{CDF:W},
using data corresponding to $8.8 fb^{-1}$ of integrated luminosity collected in proton-antiproton
collisions at a 1.96 TeV center-of-mass energy, obtain the new value of W boson mass as
\beqa
M_W=80,433.5 \pm  6.4({\rm stat}) \pm 6.9 ({\rm syst})=80,433.5\pm 9.4 {\rm MeV}/c^2~,
\eeqa
which is in significant tension with the standard model(SM) expectation which gives~\cite{SM:W}
\beqa
M_W=80,357\pm 4({\rm inputs})\pm 4 ({\rm theory}) {\rm MeV}/c^2~,
\eeqa
and the discrepancy is\cite{2205.12237}\footnote{This estimate of discrepancy can be changed by all the variations at the level of 10\%. For instance, in Refs\cite{2204.04204,2205.12237,Sci376-22-6589}, the global fit updated central values are $M_W^{(exp)} = 80.413$ GeV and $M^{(SM)}_W = 80.350$ GeV. However, it can be seen that the anomaly in the W-boson mass is certainly present.}
\beqa
\Delta M_W= 70 \pm 11 {\rm MeV}/c^2.\label{discr} 
\eeqa~
Such deviations, if got confirmed by other experiments, will strongly indicate the existence
of new physics beyond SM~\cite{Athron:muong-2,anomaly:W}. So, it is interesting to survey
what is the new constraint that the new CDF II data can impose on the new physics models
in addition to the 125 GeV Higgs.

This W mass increment has great discrepancy from the SM prediction,
which may imply the existence of new physics beyond SM.
Many attempts have been made in new physics framework, in which the anomaly usually attributes
to the deviation of oblique parameters, especially $\Delta T$\cite{STU}.

On the other hand, with the current
world-averaged result given by \cite{WAR-g-2},
the precision measurement of $a_\mu=(g-2)/2$ has been performed by the E821 experiment at
Brookhaven National Laboratory \cite{BNL-g-2},
\beq
a_\mu^{exp}= 116592091(\pm54)(\pm33)\times 10^{-11}.
\eeq
While the Standard Model (SM) prediction from the Particle Data Group gives\cite{WAR-g-2},
\beq
a_\mu^{SM}= 116591803(\pm1)(\pm42)(\pm26)\times 10^{-11},
\eeq
The difference between theory and experiment shows a $4.2\sigma$ discrepancy, hinting at tantalizing new
physics beyond the SM.

In models beyond the SM, there may exist a new confining unbroken non-abelian gauge interaction
\cite{Okun-1980,Bjorken-1979,Gupta-Quinn-1982,0805.4642,0604261,0810.1524},
in analogy to quantum chromodynamics (QCD) of strong interaction.
In general, one can assume the new color group confinement scale $\Lambda_X$ is smaller than the
QCD scale $\Lambda_{QCD}$ such that the new color degree
of freedom bears the name infracolor (IC).
We call this infracolor gluon fields igluons, and the fermions quirks\footnote{Quirks
can have their scalar partner called squirks, and they can also exist in the
same supermultiplets.}.
It was called the quirk model, which is also regarded as a certain limit of QCD with some heavy quarks
called quirks\footnote{This particle was also called "thetons"\cite{Okun-1980} or "iquark"\cite{0810.1524}},
and when QCD gets strong, the scale $\Lambda_X$ is much smaller than the quirk masses \cite{Bjorken-1979,Gupta-Quinn-1982}
with the light quarks removed from the particle list.
 Unlike the real world with light quarks, without worrying about the spontaneous chiral symmetry breaking,
this hypothetical QCD can have drastically different phenomenology.

All kinds of possible new physics beyond the SM have been searched at the Large Hadron Collider
(LHC), after the discovery of the SM Higgs boson.
%
Solutions to the gauge hierarchy problem of the SM of particle physics
such as supersymmetry and composite Higgs models usually predict a colored top partner
with mass around TeV scale. They have been challenged by the null results of LHC searches
so far. Theories of neutral naturalness \cite{1509.04284} aim to address the gauge hierarchy problem
without introducing colored states, thus relieve the tension with the LHC searches. This
class of models include folded supersymmetry \cite{0609152,0805.4667}, quirky little Higgs \cite{0812.0843},
twin Higgs \cite{0506256,1501.05310,1905.02203},
minimal neutral naturalness model \cite{1810.01882} and so on. In those models, some new $SU(N)$ gauge
symmetries are introduced in addition to the SM gauge group.

The quirk particle is charged under both the SM electroweak gauge group and the new confining $G_X$ gauge group and
has mass much larger than the confinement scale ($\Lambda_X$) of the $G_X$.
At colliders, the quirk can only be produced in pairs due to the conserved $G_X$ symmetry.
The interaction between two quirks induced by the $G_X$ gauge bosons, the infracolor force $F_s$,
will lead to non-conventional signals in the detector.
The manifestation of the quirk signal is strongly dependent on $\Lambda_X$ due to $F_s \propto \Lambda_X^2$ \cite{2002.07503,0805.4642}.
%


The quirk particle may affect large in view of its electroweak couplings with the gauge bosons and couplings to the leptons,
so we will consider the quirk contributions to the W mass increment and the $g-2$ anomaly in this paper,
which is organized as follows.
In Sec~\ref{sec-2}, we introduce the supersymmetric quirk particles and the relevant couplings.
In Sec~\ref{sec-3}, we discuss the constraints of the CDF II W boson mass data on the parameters within the quirk models.
In Sec~\ref{sec-4}, the quirk contribution to the $g-2$ anomaly will be calculated.
Sec~\ref{sec-5} gives our conclusions.

\section{The quirk particles and the relevant couplings }\label{sec-2}

The infracolor dynamics can allow the $Q\bar Q$ bound state to survive for
distances of order centimeter and preventing them from annihilation.
In some case,  production of the "squirk-antisquirk"
pair at the LHC would quickly lose their excitation energy
by bremsstrahlung and relax to the ground state of the scalar quirkonium\cite{0609152,0805.4667,1012.2072}.

In this work, we consider vector-like quirks with respect to the electroweak gauge group,
together with their scalar partners in the same supermultiplets under the framework of supersymmetry.
The new color group $G_X$ can be  $SU(2)_X$ or $SO(3)_X$ or $SU(3)_X$,
and the new fields are taken to transform in the $N=2$, $3$, or $3$ dimensional representations
respectively for these three cases.
Thus the new quirk chiral
supermultiplets containing fermion multiplet $D,~L,~S$ and their partners $\tilde D,~\tilde L,~\tilde S$ transform under $G_X \times SU(3)_c \times SU(2)_L \times U(1)_Y$\cite{1012.2072}.

$D,~L,~S$ and their partners $\tilde D,~\tilde L,~\tilde S$ may be assumed to get the masses
the same as superpotential term of the minimum SUSY models(MSSM), $\mu H_uH_d$, where  $H_u$ and $H_d$
are vector-like Higgs chiral supermultiplets in the SUSY models,
with VEVs $v_u, v_d$ ratio $\tan\beta = v_u/v_d$ and $v = \sqrt{v_u^2 + v_d^2} \approx 175$ GeV.
The non-renormalizable superpotential terms may appear as \cite{1012.2072,Kim:1983dt,Murayama:1992dj}:
\beq
W = \frac{1}{M_P^2} X \overline X  \left (
\lambda_\mu H_u H_d + \lambda_D D \overline D + \lambda_L L \overline L +
\lambda_{S_i} S_i
\overline S_i \right ),
\label{eq:axionicmasses}
\eeq
where $i=1,\ldots, n_S$ with $n_S$ SM group singlets in the same representationsof $G_X$,
and the reduced Planck mass $M_P = 2.4 \times 10^{18}$ GeV.
The fields $X, \overline X$ will get VEVs roughly of order $10^{11}$ GeV.

The vector-like mass terms in the low-energy effective superpotential can be written as\cite{1012.2072}
\beq
W = \mu H_u H_d
+ \mu_D D \overline D + \mu_L L \overline L + \mu_{S_i} S_i \overline S_i
.
\label{eq:masses}
\eeq
where $\mu, \mu_D, \mu_L, \mu_S$ can be in order of 100 GeV to 1 TeV, only if
the corresponding couplings $\lambda_\mu, \lambda_D, \lambda_L,
\lambda_S$ are not too small.

For $n_S > 0$, the new chiral supermultiplets can have Yukawa couplings
in addition to their mass terms in eq.~(\ref{eq:masses}):
\beq
W =
k_i H_u L \overline S_i + k'_i H_d \overline L S_i.
\label{eq:defineYukawas}
\eeq
On the other side, if there exists the superpotential term such as,
\vspace{-0.2cm}
\beq
W = \lambda_\ell \overline {\tilde S}\hspace{0.015in} \overline L \hspace{0.005in} \ell
\label{eq:sll}\vspace{-0.2cm}
\eeq
with $\ell$ an MSSM $SU(2)_L$ doublet lepton, we may expect it would have some influence on the muon $g-2$ discrepancy
between the experiments and the theoretical calculation.

\section{The $S,~T,~U$ parameters and $W$-mass increment} \label{sec-3}
The corrections to various electroweak precision observables can be obtained from the corresponding oblique parameters.
The new physics contributions to the W-boson mass increment can embody in the Peskin's $S,T,U$ oblique parameters~\cite{STU,STU1,STU2}
by the following\cite{STU,Spheno,W:STU},
\beqa
\Delta m_W=\f{\al m_W}{2(c_W^2-s_W^2)}(-\f{1}{2}S+c_W^2 T+\f{c_W^2-s_W^2}{4s_W^2} U)~,
\label{deltamw}
\eeqa
with
\beqa
\alpha S & = & 4s_w^2 c_w^2
               \left[ \AP0{ZZ}
                          -\frac{c_w^2-s_w^2}{s_w c_w}\AP0{Z\gamma}
                          -\AP0{\gamma\gamma}
               \right]\,,  \nonumber \\
\alpha T & = & \frac{\A0{WW}}{m_w^2} - \frac{\A0{ZZ}}{m_Z^2}\,, \\
\alpha U & = & 4s_w^2
               \left[ \AP0{WW} - c_w^2\AP0{ZZ}
                         - 2s_wc_w\AP0{Z\gamma} - s_w^2\AP0{\gamma\gamma}
               \right]\,, \nonumber
\eeqa
and $\al^{-1}(0)=137.035999084~,s_W^2=0.23126$.

 The oblique parameters $(S,~ T,~ U)$ \cite{STU}, which represent radiative corrections to the two-point functions of gauge bosons,
can describe most effects on precision measurements.
As we know, the total size of the new physics sector can be measured by the oblique parameter $S$,
while the weakisospin breaking can be measured by $T$ parameter.
The new results of $S,~ T,~ U$ can be given as \cite{2204.03796},
\beq\label{fit-stu}
S=0.06\pm 0.10, ~~T=0.11\pm 0.12,~~U=0.14 \pm 0.09.
\eeq

The most important electroweak precision constraints on quirk models comes from the
electroweak oblique parameters $S$ and $T$~\cite{STU,STU1,STU2},
and we will proceed to study the connection between the electroweak precision data with the W mass.
The model can produce main corrections to the masses of gauge bosons
via the self-energy diagrams exchanging the vector-like extra fermions.

With the Yukawa couplings $k,~k'$, the new contributions to the Peskin-Takeuchi $S,T$ observables from the new fermions
can be given as\cite{1012.2072},
\beqa
\Delta T &=& \frac{N v^4}{480 \pi s_W^2 M_W^2 M_F^2}
[13 (\hat k^4 + \hat k^{\prime 4})
+ 2 (\hat k^3 \hat k^{\prime} + \hat k \hat k^{\prime 3})
+ 18 \hat k^2 \hat k^{\prime 2} ],
\label{dt}
\\
\Delta S &=& \frac{N v^2}{30 \pi M_F^2}
[4 \hat k^2 + 4 \hat k^{\prime 2} -7 \hat k \hat k^{\prime} ].
\label{ds}
\eeqa
where $\hat k = k \sin\beta$ and $\hat k' = k' \cos\beta$ and $v \approx
175$ GeV,

In our analysis, we will perform a global fit to the predictions of $S,~T$ parameters in profiled $1\sigma$ favoured regions.
We scan $m_F$, $tan\beta$, $k$ and $k'$ parameters in the following ranges:
\vspace{-0.65cm}
\begin{align}
100 {\rm ~GeV} \leq m_F \leq 1100 {\rm ~GeV},~~ 1\leq tan\beta \leq 50, ~~~ 0.01  < k,k' < 1.
\label{epot2}
\end{align}
%
%
\begin{figure*}[]
   \centering
    \includegraphics[width=0.482\textwidth]{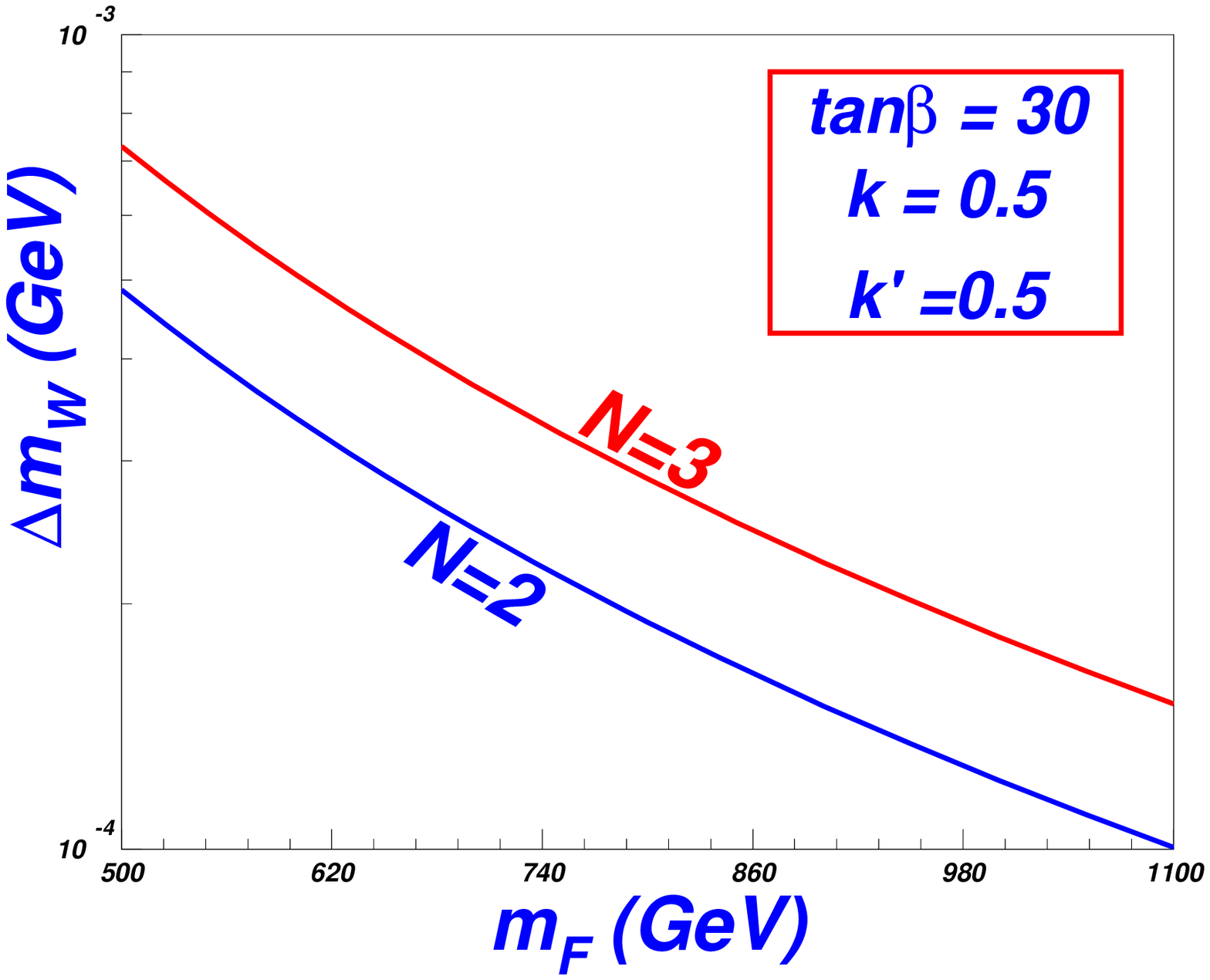} 
    \includegraphics[width=0.482\textwidth]{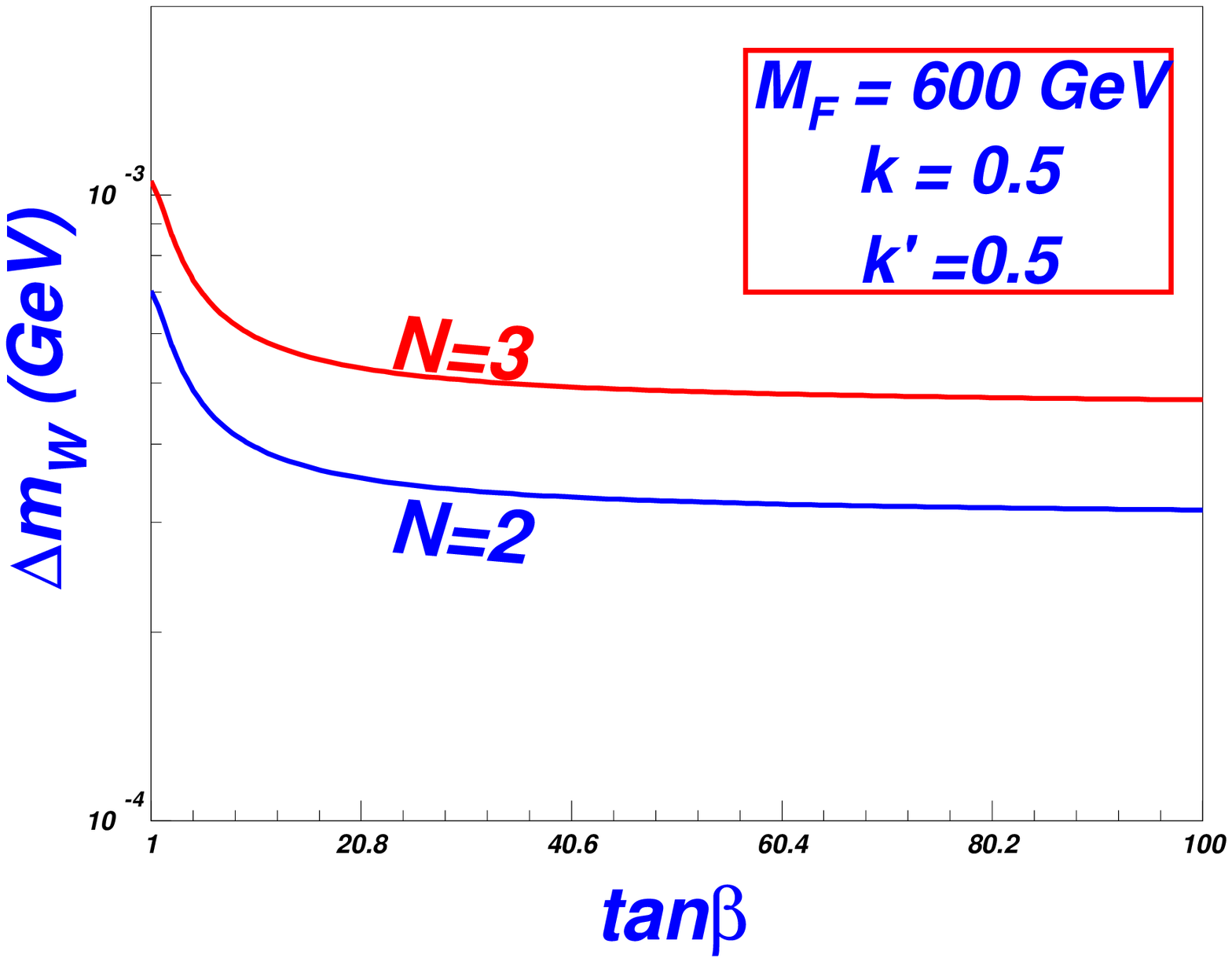}
   \caption{The W mass increment varies with varying $m_F$ and $tan\beta$. }
 \label{fig1}
\end{figure*}
\begin{figure}[]
    \centering 
    \includegraphics[width=0.482\textwidth]{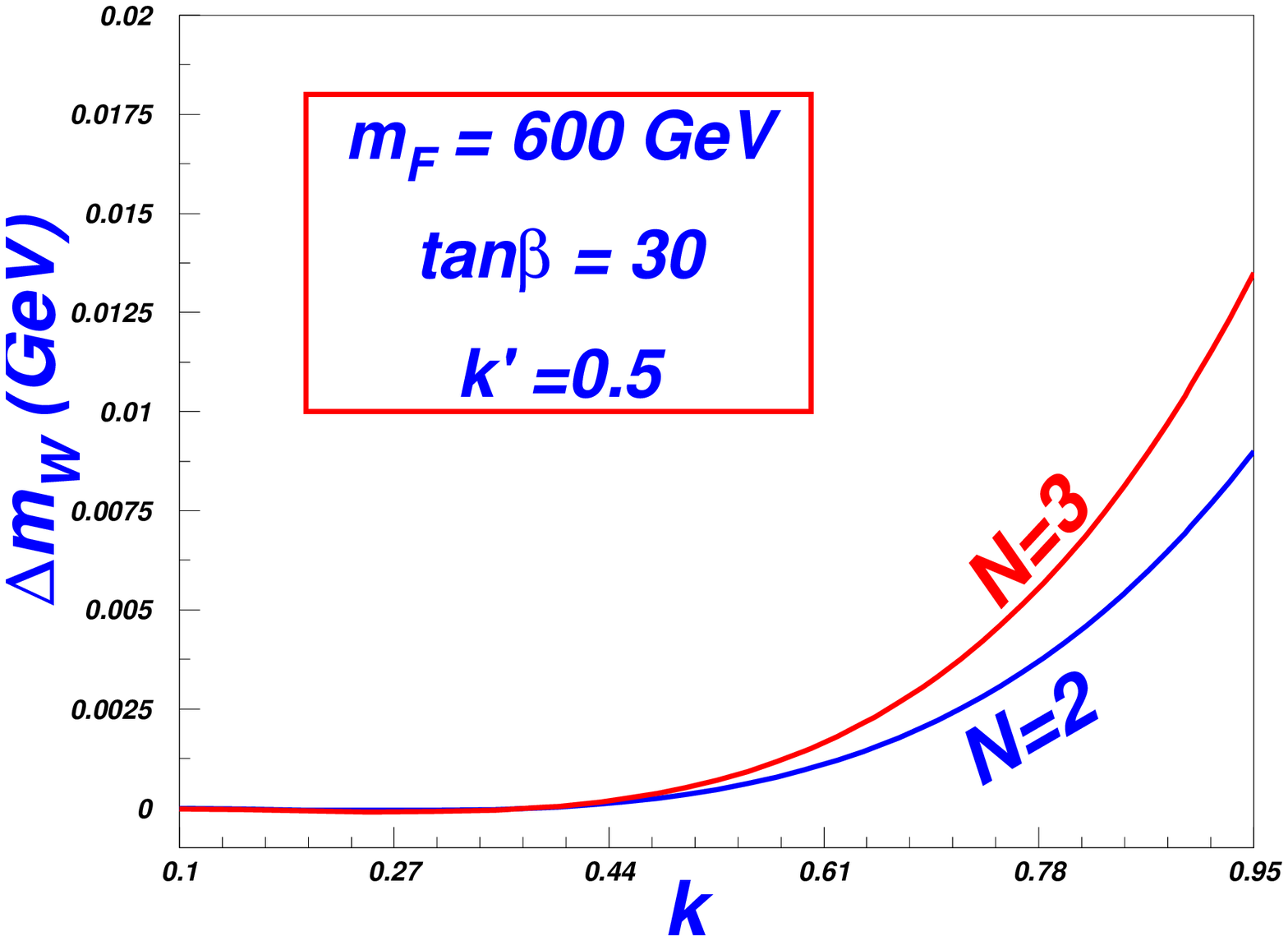}
    \includegraphics[width=0.482\textwidth]{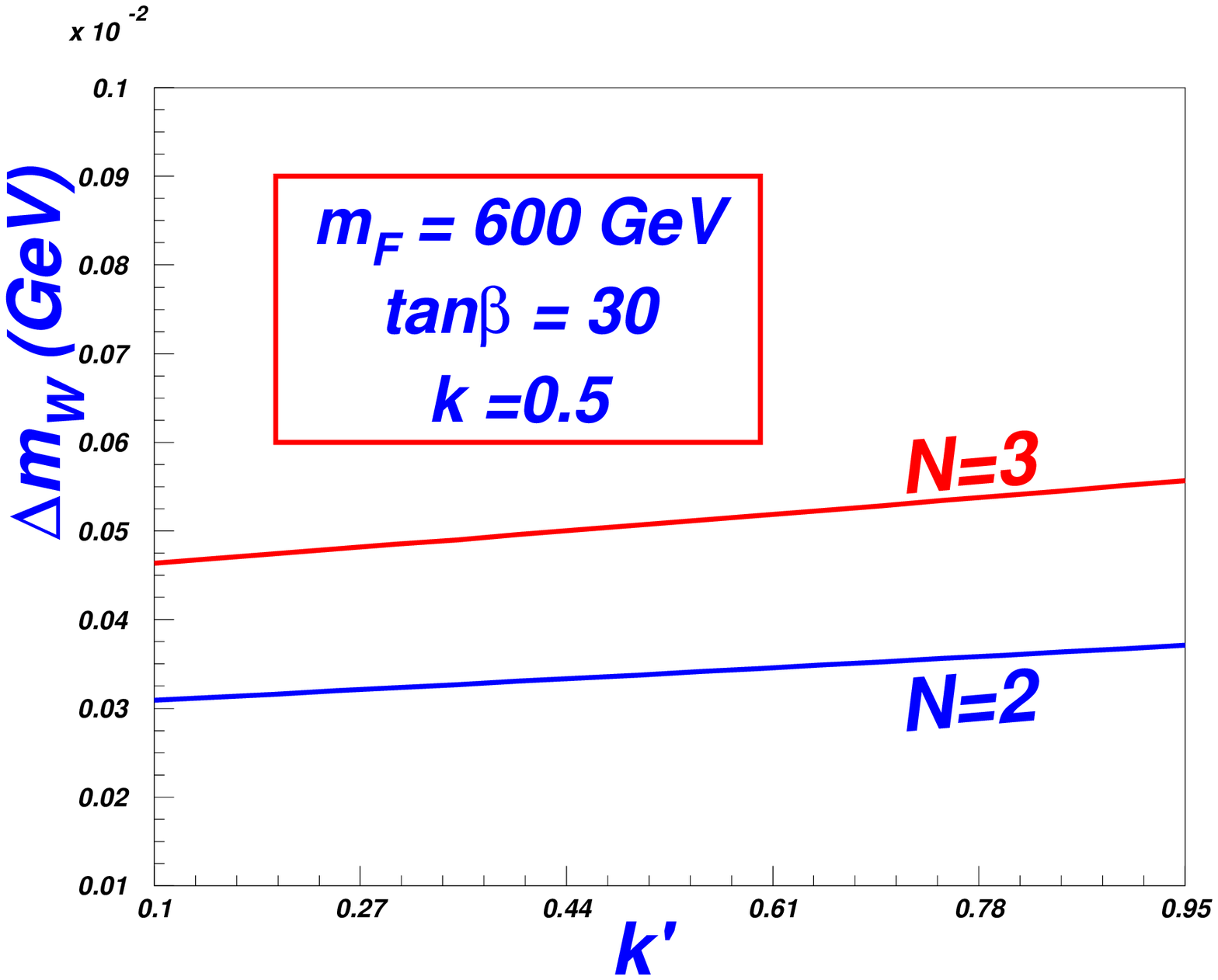}
    \caption{Same as Fig.\ref{fig1}, but with varying $k$ and $k'$. }
 \label{fig2}
\end{figure} 

In Fig. \ref{fig1} and Fig. \ref{fig2}, we show the W mass increment varies with the ratio $\tan\beta$, Fermion massws $m_F$, and $k,~k'$, which are in the range of (1-50), ($100-1100$ GeV), and $( 0.01-1$), respectively, with the fixed parameters shown in the figures.
From the two figures, we can see that the W mass increment decreases monotonously with increasing $m_F,~tan\beta$, while increases monotonously with increasing $N,~k,~k'$.
The dependence of $N$ and $m_F$ of the W mass increment is obvious.

But when $tan\beta$ gets larger, the influence will be smaller and smaller,
That is because, from Eq.(\ref{dt}) and (\ref{ds}), the ratio $tan\beta$
appears in $\Delta S$ and $\Delta T$ in the form of $sin\beta$ and $cos\beta$, the former of which is approaching to the maximum value $1$ and the latter, to the minimum value 0, with the larger and larger $tan\beta$ values.

We can also see that the  effect from the couplings $k$ and $k'$ is not
equivalent in Fig. \ref{fig2} and contribution range from $k$ can span from negative to positive $10^{-2}$ GeV, while that from $k'$, just in $10^{-3}$ GeV. The reason of the in insensitivity of $k'$ is that in the oblique parameters Eq.(\ref{dt}) and (\ref{ds}), $k'$ is always multiplied by $cos\beta$, which is too small when $tan\beta=30$ we choose, just as shown in Fig. \ref{fig1} and Fig. \ref{fig2}.
So it is necessary to consider the contributions in the whole parameter space.

By the way, in most parameter space, the W mass increment is in the experimental limit, so we may try to constrain the parameters according to the bound of the W mass discrepancy between the experiments and the SM prediction in Eq.(\ref{discr}).

%
\begin{figure}[]
   \centering
    \includegraphics[width=0.482\textwidth]{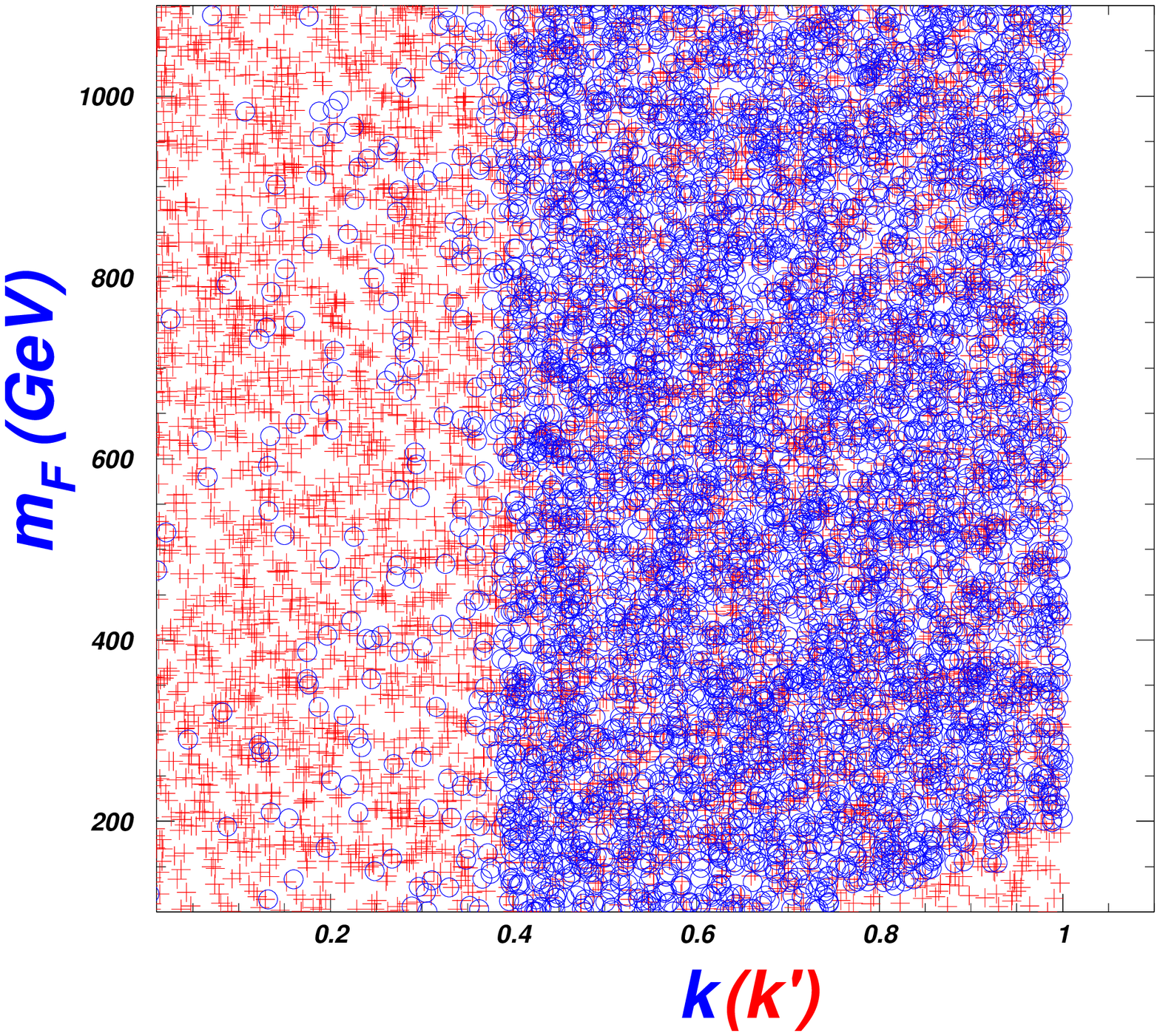}
    \includegraphics[width=0.482\textwidth]{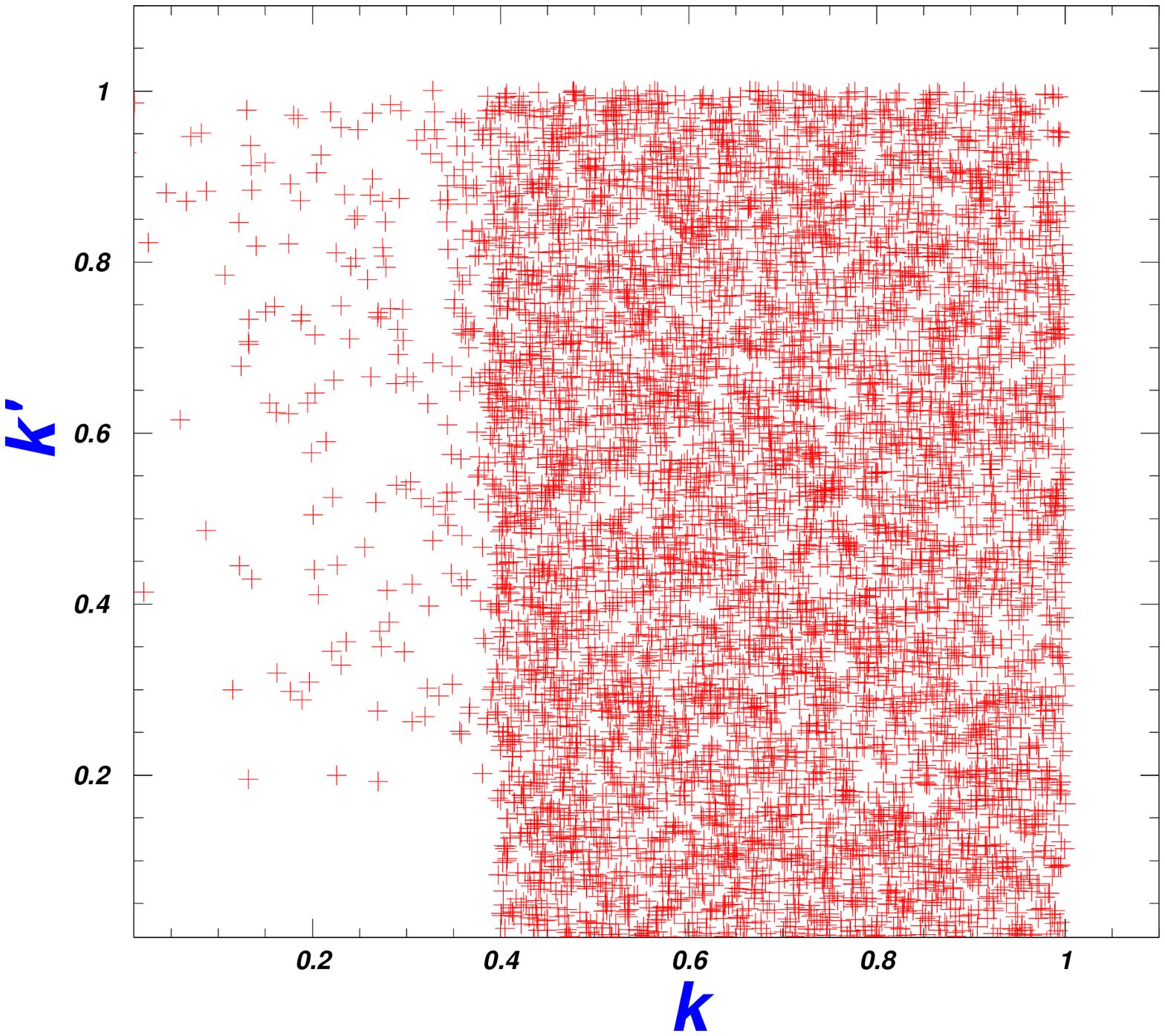}
    \includegraphics[width=0.582\textwidth]{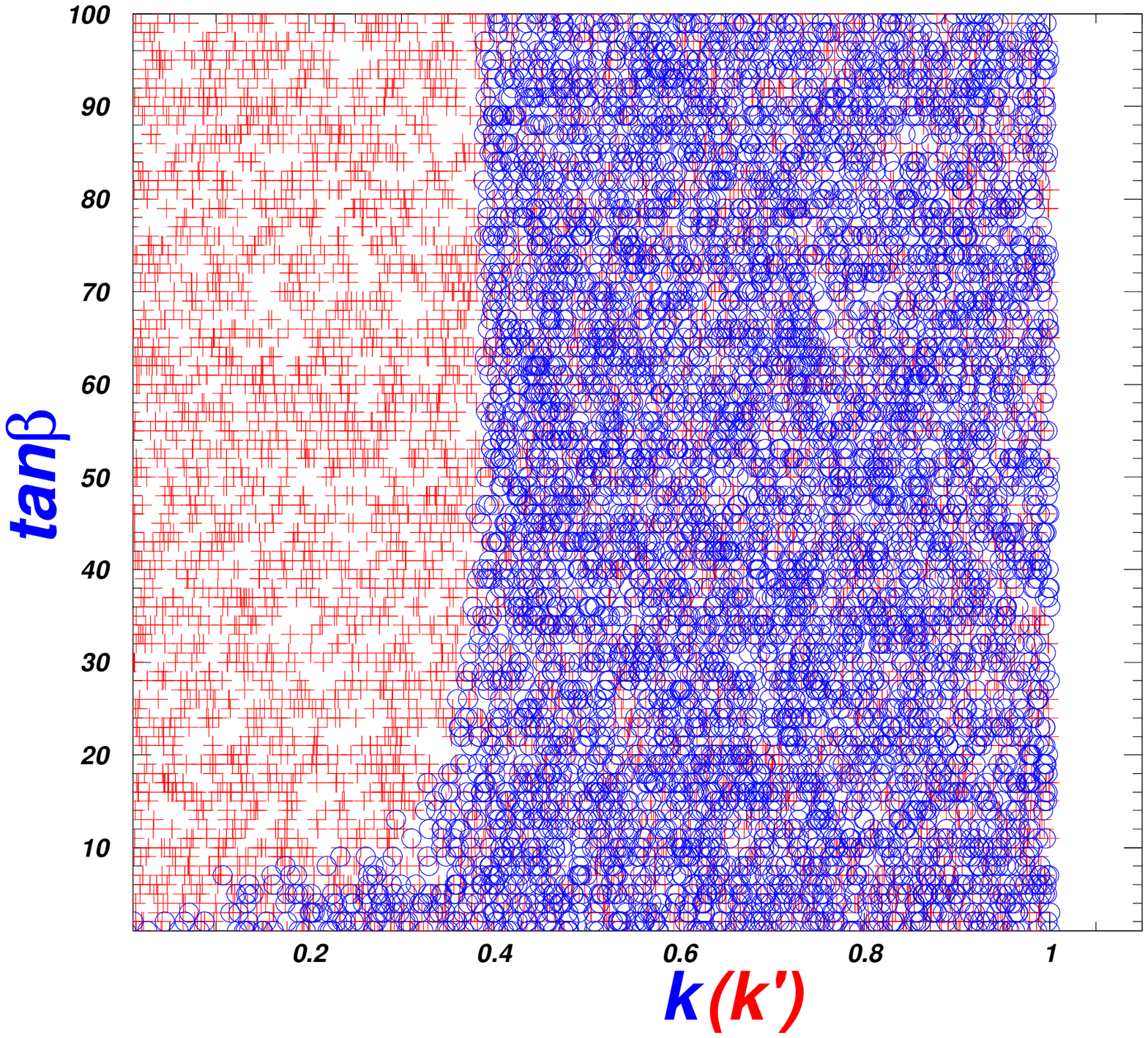}\vspace{-0.4cm}
 \caption{The samples explaining the CDF II results of the $W$-mass within $1\sigma$ range. }
 \label{fig3}
\end{figure} 
 The above constraints on the W increment mass from the parameters $N,~m_F,~tan\beta,~k,~k'$ are obtained independently.
 In Fig. \ref{fig3} we consider the joint effect by scanning the allowed points possible to exist for the mass increment in the $1\sigma$ range of the experimental bound. We set $10000$ scanned random points, and $6149$ points meet the constraints in the scanning.
Note that we have fixed $N=2$ in Fig. \ref{fig3} since it does not exert much influence on the results.

From the first diagram of Fig. \ref{fig3}, we can see that
there are almost no constraints on $m_F$, and it can be any value selected in the scanning.
That is because in our parameter spaces, within the whole possible value of the fermion mass $m_F$, 
it can be seen from Eqs. (\ref{dt}), (\ref{ds}), the coefficient of $\Delta T$ is much larger than that of $\Delta S$ and the contributions of the following terms of the twos are of the same size, so the contribution of $\Delta T$ is primary.
On the other hand, the coefficient is much smaller than the subsequent terms.
 Then from Eq. (\ref{deltamw}) we see that the factor that the coefficient of $\Delta T$ is larger than that of $\Delta S$, ensures that they together contribute positively to $\Delta m_W$, and the coefficients are smaller than the subsequent terms, while $m_F$ is a part of the coefficients. Therefore,
 as long as the result is positive, $m_F$ hardly affects $\Delta m_W$, that is to say, $m_F$ is not restrained by the CDF data.

From the first diagram of Fig. \ref{fig3}, we can also see that the contributions from $k$ and $k'$ are not the same, just same as the discussed above, and $k$ is bound as $k>0.4$, while $k'$ ranges the whole space. The reason for the insensitivity of $k'$ is that $k'$ is always multiplied by the factor $cos\beta$, which is small with large $\tan\beta$, which starts from $1$.

 The third diagram of Fig. \ref{fig3} shows that the constraints on $\tan\beta$ is also quite weak, which can also be seen in the right diagram of Fig.\ref{fig1},
 that is because since $k$ and $sin\beta$, $k'$ and $cos\beta$ appear together, and the relation $sin^2\beta+cos^2\beta=1$ will finally decrease the contribution with
 increasing $tan\beta$.

 Thus we can conclude that in the most of the parameter space, the parameters in supersymmetric quirk models can account for the CDF data of the W mass increment,
and only the constraints on $k$ is obvious, $k\gtrsim 0.4$.

\section{The $g-2$ anomaly of the new couplings} \label{sec-4}

\begin{figure*}[htb]
   \centering
  \includegraphics[scale=1.5]{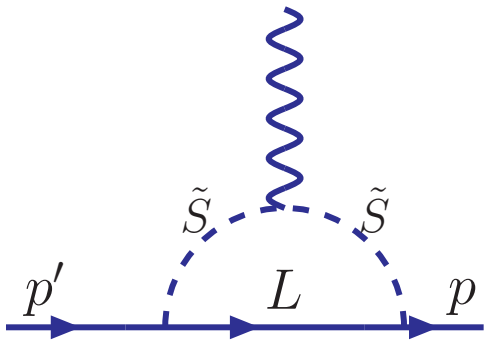}
\vspace{-0.3cm}
\caption{ The one-loop contributions to $a_\mu$ in quirks models.}
\label{fig4}
\end{figure*}
 In quirk models, the muon $g-2$ contributions are mainly obtained via the
one-loop diagrams induced by the couplings shown in Eq.(\ref{eq:sll}) as in Fig.\ref{fig4}.
Note that two-loop Barr-Zee diagrams disappear since there is no mixing between SM gauge bosons of the new
quirk particles \cite{1012.2072} and the $m_\mu^2/m_S^2$ suppression of the diagram containing such two couplings\cite{1502.04199}.
 \begin{figure}[]%
    \centering
  \includegraphics[width=0.582\textwidth]{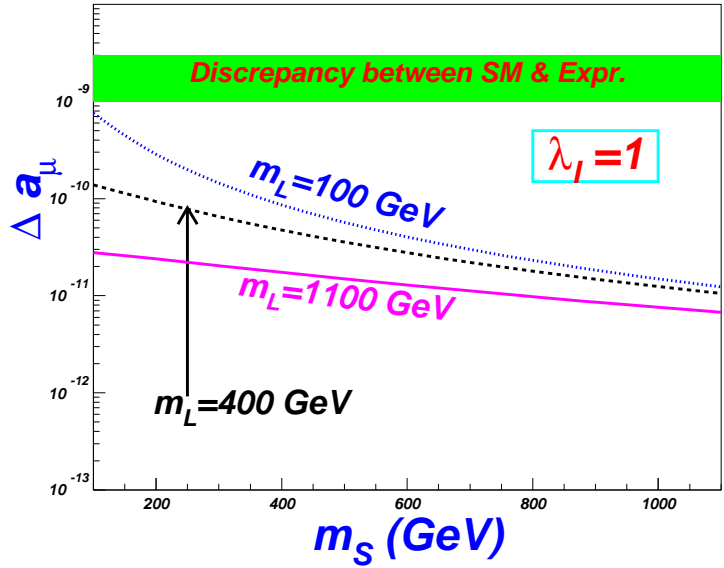}\vspace{-0.3cm}
 \caption{ The one-loop muon anomalous magnetic moment varying with $m_S$ for $\lambda_\ell^2=1$ and
$m_L=100,~400,~1100$GeV, respectively.
The green shadow area is the discrepancy between the SM
and the measurement for the anomalous magnetic moment $\Delta a_\mu$.}
 \label{fig5}
  \end{figure}

The one-loop contribution can be written as\cite{1loop-deltaamu,g-2-th}:
 \beq
    \Delta a_\mu^{\mbox{$\scriptscriptstyle{\rm qm}$}}({\rm 1-loop})_H^{(b)} =
   \lambda_\ell^2 \frac{m_\mu^2}{16\pi^2 } \int_0^1 dx \frac{x^3-x^2}{m_S^2 x+ m_L^2(1-x)},
\label{1loopH}
\eeq
where $\lambda_\ell^2$ is the coupling shown in Eq.(\ref{eq:sll}), on which the one-loop moment magnetic can be realized as Fig.\ref{fig4}.

Since $\tilde S$($L$) is the scalar(fermion) component of the supermultiplet, their masses should be
in the same level as that of the fermion $F$, so we take them changing also in the range of $100-1100$ GeV.
We scan coupling $\lambda_\ell^2$ from $0$ to $1$.

Fig.\ref{fig5} shows that the contributions of scalar and fermion from the supermultiplet
at one-loop level, and we find that the contribution is quite small,
 about $\sim 10^{-10}$, which can not explain the the discrepancy
 between the experiments and the theoretical prediction.
The situation is not surprising, since it has been pointed out that only if the 
scalar masses are very small, such as several GeV, the contribution may be large\cite{Ellwange-fwang},
but our choice for the new particle masses is larger than $100$ GeV. 

Hence, at the one-loop level, it is difficult to fill the gap between the experiments and the theoretical prediction in the supersymmetric quirk models.
Therefore, with the missing two-loop Barr-Zee diagram, we can conclude that
the supersymmetric quirk models can not account for the muon $g-2$ anomaly.


\section{\label{sec-5}Conclusions}

In this paper,  we firstly show the W mass increment varies with the parameters $\tan\beta$, $m_F$, and $k,~k'$ with different new color group representations $N$,
and we find the dependence on the parameters of the W mass increment is obvious.
We then scan the allowed points possible to exist for the mass increment
in the $1\sigma$ range of the experimental bound
and find that there are almost no constraints on $m_F$,
and the contributions from $k$ and $k'$ are not the same, with $k>0.4$, while $k'$ ranges the whole space.
 and the constraints on $\tan\beta$ is also quite weak.
In a word, in the most of the parameter space,
supersymmetric quirk models can account for the CDF data of the W mass increment,
and only the constraints on $k$ is obvious, $k\gtrsim 0.4$.

We also calculate the contribution to the muon $g-2$ anomaly at the one-loop level,
and find that it is difficult to account for the gap between the experiments and
the theoretical prediction in the supersymmetric quirk models.

\section*{Acknowledgment}
 This work was supported by the National Natural Science Foundation of China(NSFC)
under grant 12075213, 
 by the Fundamental Research Cultivation Fund for Young Teachers of Zhengzhou University(JC202041040)
 and the Academic Improvement Project of Zhengzhou University.



\begin{thebibliography}{99}


\bibitem{CDF:W} CDF Collaboration et al., Science 376, 170-176 (2022).
\bibitem{SM:W} P. A. Zyla et al., Prog. Theor. Exp. Phys. 2020, 083C01 (2020).
\bi{2205.12237}S. Afonin, W-boson mass anomaly as a manifestation of spontaneously broken additional SU(2) global symmetry on a new fundamental scale, Universe 8 (2022) 627, arXiv:2205.12237.
\bi{2204.04204}J. de Blas, M. Pierini, L. Reina, L. Silvestrini,
Impact of the Recent Measurements of the Top-Quark and W-Boson Masses on Electroweak Precision Fits, Phys. Rev. Lett. 129 (2022) 27, 271801, arXiv:2204.04204.
\bi{Sci376-22-6589}
T. Aaltonen, et al. [CDF Collaboration], Science 376 (2022) 6589.

\bibitem{Athron:muong-2} P.~Athron, C.~Bal\'azs, D.~H.~J.~Jacob, W.~Kotlarski, D.~St\"ockinger and H.~St\"ockinger-Kim, JHEP \textbf{09} (2021), 080
doi:10.1007/JHEP09(2021)080
[arXiv:2104.03691 [hep-ph]].
\bibitem{anomaly:W} E.~Bagnaschi, M.~Chakraborti, S.~Heinemeyer, I.~Saha and G.~Weiglein,
[arXiv:2203.15710 [hep-ph]].
\bibitem{STU}  M.E. Peskin, T. Takeuchi, Phys. Rev. Lett. 65 (1990) 964;
               M.E. Peskin, T. Takeuchi, Phys. Rev. D 46 (1992) 381.



\bibitem{WAR-g-2} C. Patrignani et al. (Particle Data Group), Chin. Phys.
C40, 100001 (2016).
\bibitem{BNL-g-2} G. W. Bennett et al. (Muon g-2), Phys. Rev. D73,
072003 (2006), arXiv:hep-ex/0602035.


\bibitem{Okun-1980}[1] L. B. Okun, JETP Lett. 31, 144 (1980) [Pisma Zh. Eksp. Teor. Fiz. 31, 156 (1979)]; Nucl.
Phys. B 173, 1 (1980).
\bibitem{Bjorken-1979}[2] J. D. Bjorken, SLAC-PUB-2372 (1979), in Quantum Chromodynamics, proceedings of the
SLAC Summer Institute on Particle Physics, Stanford, California, 1979, edited by Anne
Mosher (SLAC, Stanford, 1980).
\bibitem{Gupta-Quinn-1982}[3] S. Gupta and H. R. Quinn, Phys. Rev. D 25, 838 (1982).
\bibitem{0805.4642}[4] J. Kang and M. A. Luty, arXiv:0805.4642 [hep-ph].
     \bi{0604261} M. J. Strassler and K. M. Zurek, Phys. Lett. B 651, 374 (2007), [arXiv:hep-ph/0604261].
\bi{0810.1524}	
Kingman Cheung, Wai-Yee Keung, Tzu-Chiang Yuan, Nucl.Phys.B 811, (2009) 274, arXiv: 0810.1524.

\bi{1509.04284}[1] D. Curtin and P. Saraswat, 
     Phys. Rev. D93, (2016), no. 5 055044, [arXiv:1509.04284].

\bi{0609152}[2] G. Burdman, Z. Chacko, H.-S. Goh, and R. Harnik, 
          JHEP 02 (2007) 009, [hep-ph/0609152].
\bi{0805.4667}[3] G. Burdman, Z. Chacko, H.-S. Goh, R. Harnik, and C. A. Krenke,
         Phys. Rev. D78 (2008) 075028, [arXiv:0805.4667].
\bi{0812.0843}[4] H. Cai, H.-C. Cheng, and J. Terning, 
         JHEP 05 (2009) 045, [arXiv:0812.0843].
\bi{0506256}[5] Z. Chacko, H.-S. Goh, and R. Harnik,
         Phys. Rev. Lett. 96 (2006) 231802, [hep-ph/0506256].
\bi{1501.05310}[6] N. Craig, A. Katz, M. Strassler, and R. Sundrum,
         JHEP 07 (2015) 105, [arXiv:1501.05310].
\bi{1905.02203}[7] J. Serra, S. Stelzl, R. Torre, and A. Weiler, 
         JHEP 10 (2019) 060, [arXiv:1905.02203].
\bi{1810.01882}[8] L.-X. Xu, J.-H. Yu, and S.-H. Zhu, 
         Phys. Rev. D 101, 095014 (2020), arXiv:1810.01882.
\bi{2002.07503}Jinmian Li, Tianjun Li, Junle Pei, Wenxing Zhang,
         Euro. Phys. J. C 80, 651 (2020), arXiv:2002.07503.


\bi{1012.2072}Stephen P. Martin, 
             Phys. Rev. D 83, (2011) 035019, arXiv: 1012.2072.

\bibitem{Kim:1983dt}  J.E.~Kim and H.P.~Nilles,  
  Phys.\ Lett.\  B {\bf 138}, 150 (1984).
\bibitem{Murayama:1992dj} H.~Murayama, H.~Suzuki and T.~Yanagida, Phys.\ Lett.\  B {\bf 291}, 418 (1992).


\bibitem{STU1} W.J. Marciano, J.L. Rosner, Phys. Rev. Lett. 65 (1990) 2963;
            W.J. Marciano, J.L. Rosner, Phys. Rev. Lett. 68 (1992) 898, Erratum.
\bibitem{STU2} G. Altarelli, R. Barbieri, Phys. Lett. B 253 (1991) 161.
\bibitem{Spheno}  W. Porod, Comput. Phys. Commun. 153 (2003) 275 [arXiv:hep-ph/0301101];\\
                  W. Porod and F. Staub, Comput. Phys. Commun. 183 (2012) 2458 [arXiv:1104.1573].


\bibitem{W:STU} R. Boughezal, J.B. Tausk, J.J. van der Bij, Nuclear Physics B 725 (2005) 3-14.
\bibitem{2204.03796}C.-T. Lu, L. Wu, Y. Wu, B. Zhu, arXiv:2204.03796.
\bibitem{top-bottom-seesaw} H.-C. Cheng, B. A. Dobrescu, J. Gu, JHEP08(2014)095,  arXiv: 1311.5928;
C. Balazs, T. Li, F. Wang and J. M. Yang, JHEP 1301, 186 (2013), arXiv:1208.3767.

\bi{1502.04199} Victor Ilisie, New Barr-Zee contributions to (g-2)mu in two-Higgs-doublet models,
 JHEP 04, (2015) 077, arXiv:1502.04199.

 \bibitem{1loop-deltaamu}
J. P. Leveille, Nucl. Phys. B 137, 63 (1978);
S. R. Moore, K. Whisnant, and Bing-Lin Young, Phys. Rev. D 31, (1985) 105;
Farinaldo S. Queiroz, William Shepherd, Phys. Rev. D 89 (2014) 095024, arXiv:1403.2309.
\bibitem{g-2-th} Guo-Li Liu,Ping Zhou,
The Contribution of Charged Bosons with Right-Handed Neutrinos to the Muon g-2 Anomaly in the Twin Higgs Models,
Universe 8 (2022) 12, 654, arXiv:2101.00607.

\bi{Ellwange-fwang}See e.g,
Domingo, F.; Ellwanger, U. Constraints from the Muon g-2 on the Parameter Space of the
NMSSM. J. High Energy Phys. 2008, 2008, 79.
https://doi.org/10.1088/1126-6708/2008/07/079 \\
Explaining the Muon g-2 Anomaly in Deflected AMSB for NMSSM, Li-Jun Jia, Zhuang Li, Fei Wang,
Universe 2023, 9, 214, arXiv:2305.04623.


\end{thebibliography}
\end{document}